# Ultrahigh-Resolution Fiber-Optic Sensing Using a High-Finesse, Meter-Long Fiber Fabry-Perot Resonator

Nabil Md Rakinul Hoque and Lingze Duan, *Senior Member, IEEE, Senior Member, OSA*

*Abstract*— Ultrahigh-resolution fiber-optic sensing has been demonstrated with a meter-long, high-finesse fiber Fabry-Perot interferometer (FFPI). The main technical challenge of large, environment-induced resonance frequency drift is addressed by locking the interrogation laser to a similar meter-long FFPI, which, along with the FFPI sensor, is thermally and mechanically isolated from the ambient. A nominal, noise-limited strain resolution of 800 $f\varepsilon$ /√Hz has been achieved within 1–100 Hz. Strain resolution further improves to 75 $f\varepsilon$ /√Hz at 1 kHz, 60 $f\varepsilon$ /√Hz at 2 kHz and 40 $f\varepsilon$ /√Hz at 23 kHz, demonstrating comparable or even better resolutions than proven techniques such as $\pi$-phase-shifted and slow-light fiber Bragg gratings. Limitations of the current system are analyzed and improvement strategies are presented. The work lays out a feasible path toward ultrahigh-resolution fiber-optic sensing based on long FFPIs.

## I. INTRODUCTION

Passive fiber-optic sensors such as fiber Bragg gratings (FBG) and fiber Fabry-Perot Interferometers (FFPI) have demonstrated tremendous potential to achieve ultrahigh-resolution (UHR) optical sensing [1], [2]. Most high-resolution fiber-optic sensors share similar working principles. When external disturbance (e.g., longitudinal strain or temperature fluctuation) is applied, internal parameters such as grating period, cavity length and refractive index are subject to a change. This in turn triggers a detuning of the spectral features associated with these parameters [1]. For an FBG sensor, the Bragg reflection peak is the characteristic spectral marker [3], whereas for an FFPI, the resonance transmission peak typically serves as the indicator for spectral changes [4]. In either case, the chief goal for improving sensing resolution is to create spectral features as narrow as possible. In essence, this can be regarded as making wavelength discriminators with the highest possible wavelength selectivity [5].

A widely used metric to gauge the resolving power of a fiber-optic sensor is *strain resolution*, which is defined as the minimum change of length (per unit of the total length) a sensor is able to resolve. Over the years, a number of techniques have been developed to enhance the strain resolutions of passive fiber-optic sensors [6]-[35]. Two notable examples are phase-shifted FBG and slow-light FBG. $\pi$-phase-shifted FBG utilizes a $\lambda/4$ gap in between two identical Bragg gratings to create a very narrow peak in the middle of the reflection spectrum [6], [7]. This sharp spectral feature has proved to be effective in achieving ultrahigh strain resolution [8]-[12]. In particular, a recent report by Liu *et al.* has demonstrated a strain resolution of 140 $f\varepsilon$/√Hz at 1 kHz by locking a laser to the resonant peak of a $\pi$-phase-shifted FBG [12]. Meanwhile, ultrahigh strain resolution has also been realized with extremely narrow slow-light resonance peaks [13]-[18]. These peaks exist near both edges of the band gap of an FBG under the conditions of strong grating-index modulation, low internal loss, optimized apodization of index profile, and a suitable length [13]-[16]. They are referred to as "slow-light" resonances because of the large group delay in the vicinity of these narrow resonances [17]. It has been shown that such ultra-narrow spectral features can lead to ultrahigh sensing resolutions. For example, by probing slow-light resonances, Skolianos *et al.* has demonstrated strain resolutions of 30 $f\varepsilon$/√Hz at 30 kHz and 110 $f\varepsilon$/√Hz at 2 kHz [18].

Generating ultra-narrow resonance peaks from FFPI is another way to achieve UHR optical sensing [19]-[26]. In order to minimize the resonance linewidth of a Fabry-Perot (FP) cavity, *high mirror reflectivity* (high finesse) and *long cavity* (small free spectral range or FSR) are generally desired [27]. However, unique challenges arise in the improvement of both factors in FFPI. First, the end reflectors of most FFPIs are made of FBGs, whose peak reflectivity can typically reach up to 99%. As a result, most of the FFPI sensors reported so far feature finesses below 300 [20]-[22]. Secondly, despite clear superiority of FFPI over free-space FP in compactness and cost [28],[29], the development of UHR FFPI sensors has been hindered by the technical challenges of locking laser frequencies to long FFPI cavities. This is because the intracavity beam in an FFPI takes multiple passes inside a dielectric medium, which is much more susceptible to external perturbations (e.g., temperature fluctuation and mechanical stress) and fundamental thermodynamic noise than vacuum [30]-[32]. As a result, the resonance peaks of an FFPI usually

This work was supported in part by the National Science Foundation under Grants ECCS-1254902 and ECCS-1606836 and in part by the National Aeronautics and Space Administration under Grant 80NSSC19M0033.

N. M. R. Hoque and L. Duan are with the Department of Physics, the University of Alabama in Huntsville, Huntsville, AL 35899 USA (e-mail: nh0014@uah.edu).

exhibit large jitter and drift, making high-resolution laser interrogation very difficult [32]. Most of the FFPI-based sensors reported so far are about or less than 20 cm [19]-[24]. A notable work by Gagliardi *et al*. has demonstrated strain resolutions of 220 $f\varepsilon/\sqrt{Hz}$ at 1.5 kHz and 350 $f\varepsilon/\sqrt{Hz}$ at 5 Hz, which are accomplished by interrogating a 13-cm FFPI with a diode laser frequency-locked to an optical frequency comb (OFC) [34]. The same group has also reported the use of a 50-cm FFPI to achieve a 60 $p\varepsilon/\sqrt{Hz}$ resolution at about 900 Hz [25].

Despite these prior efforts, UHR optical sensing based on long FFPI remains to be an uncharted area. Meanwhile, it has recently been shown that laser-frequency referencing to long fiber cavities plays a vital role in the probing of fundamental thermomechanical fluctuations inherent in optical fibers [26],[35]. In this letter, we present what we believe as the first experimental realization of UHR fiber-optic sensor based on a *meter-long*, *high-finesse* (~1000) FFPI. In doing so, we have also demonstrated direct frequency locking between a diode laser and a 1-m fiber FP cavity. With only off-the-shelf components and without any additional laser-frequency stabilization, we have achieved nominal strain resolutions of approximately 800 $f\varepsilon/\sqrt{Hz}$ within 1–100 Hz and as low as 40 $f\varepsilon/\sqrt{Hz}$ at higher frequencies.

## II. EXPERIMENTAL METHOD

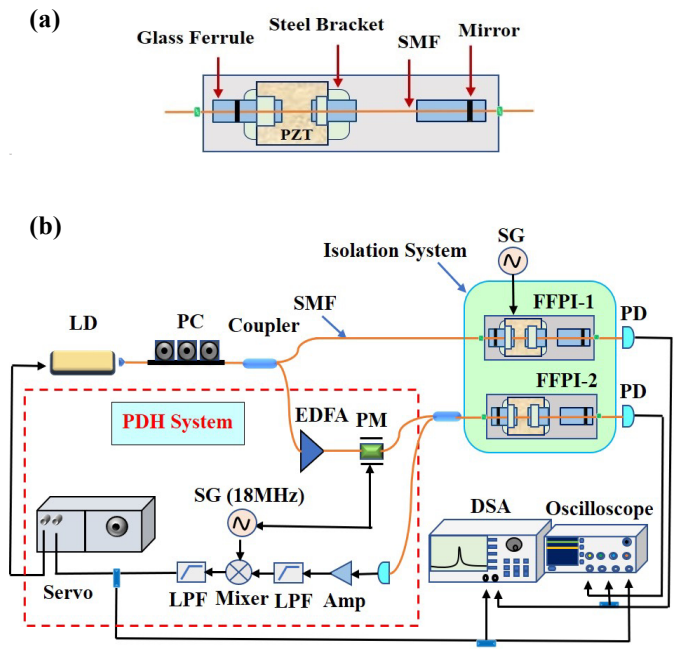

Fig. 1. (a) Schematic of the fiber Fabry-Perot interferometer (FFPI) sensor with a 1-m cavity length. (b) A layout of the overall experimental setup (for both laser-cavity locking and strain measurement). Amp: rf amplifier; LD: laser diode; LPF: low-pass filter; PD: photodetector; PM: phase modulator; SG: signal generator; SMF: single-mode fiber.

The FFPI sensor used in this research is a commercial fiber FP scanning interferometer (Micron Optics, FFP-SI). The structure of the FFPI is shown in Fig. 1(a) [36]. The interferometer body is a 1-m long single-mode fiber. Its two ends are coated with highly reflective multilayer dielectric mirrors to form an optical cavity. The cavity is specified to have a finesse of 1000. Our independent characterization indicates an actual finesse of 902, with a free-spectral range of 105 MHz and an average full-width-at-half-maximum (FWHM) linewidth of 116 kHz. The FFPI is equipped with a piezoelectric (PZT) stretcher, allowing the cavity length and the resonance peaks to be tuned.

In order to interrogate an FFPI sensor, the probing laser must be able to *reliably* track the motion of a resonance peak. This is especially difficult for a high-finesse, long fiber cavity because of the extremely narrow resonances (~100 kHz in our case). We take on this challenge from two aspects: (*i*) minimizing the random jitter and drift of the resonance peaks due to environment-induced fluctuations, and (*ii*) letting the interrogating laser *track* the large but slow drift of the FFPI so the laser can stay on resonance with the cavity for an extended period. To tackle the first task, we thermally and acoustically isolate the sensor from ambient by sealing it inside a fiberglass box, whose interior walls are lined with sound absorbing foams. The box is then placed on a passive vibration-isolation platform (Minus K, BM-1) to eliminate the influence from floor vibration. To accomplish the second goal, we introduce a reference FFPI, which is identical to the sensor FFPI, and package both FFPIs in the same isolation box. We then frequency-lock the interrogation laser to the reference FFPI. Since both FFPIs experience the same environmental perturbations, their resonance peaks have similar drift. Thus, locking the laser frequency to the reference FFPI enables the laser to track the fluctuations of the sensor resonances.

Fig. 1(b) shows a diagram of the entire experimental setup. The interrogation laser is an off-the-shelf, single-frequency diode laser (RIO, Orion), operating at 1557.4 nm with a 6-kHz linewidth. Using such a low-noise laser minimizes the impact of laser noise on the overall system noise floor. After passing through a polarization controller (PC), the laser output is split into two paths by a 90:10 fiber coupler. The majority of the optical power is coupled into the FFPI sensor (FFPI-1 in Fig. 1), while a small portion of the laser power is directed toward the reference fiber cavity (FFPI-2). The reference arm consists of a typical Pound-Drever-Hall (PDH) frequency locking system [37], preceded by an erbium-doped fiber amplifier (EDFA) for power adjustment. The PDH system operates at 18 MHz and the effective servo bandwidth is 10 Hz. The servo output is used to control the driving current of the diode laser. The particularly low servo bandwidth is intended to allow the laser frequency to track the slow drift of the FFPI resonance without following the high-frequency jitter of the resonance. The optical transmissions through both FFPIs are detected by two identical photodetectors (Thorlabs, DET20C). The detector outputs are monitored with both an oscilloscope and a fast-Fourier-transform dynamic signal analyzer (DSA) (Stanford Research Systems, SR785). Also being monitored on the oscilloscope and the DSA are the PDH error signal as well as the servo output.

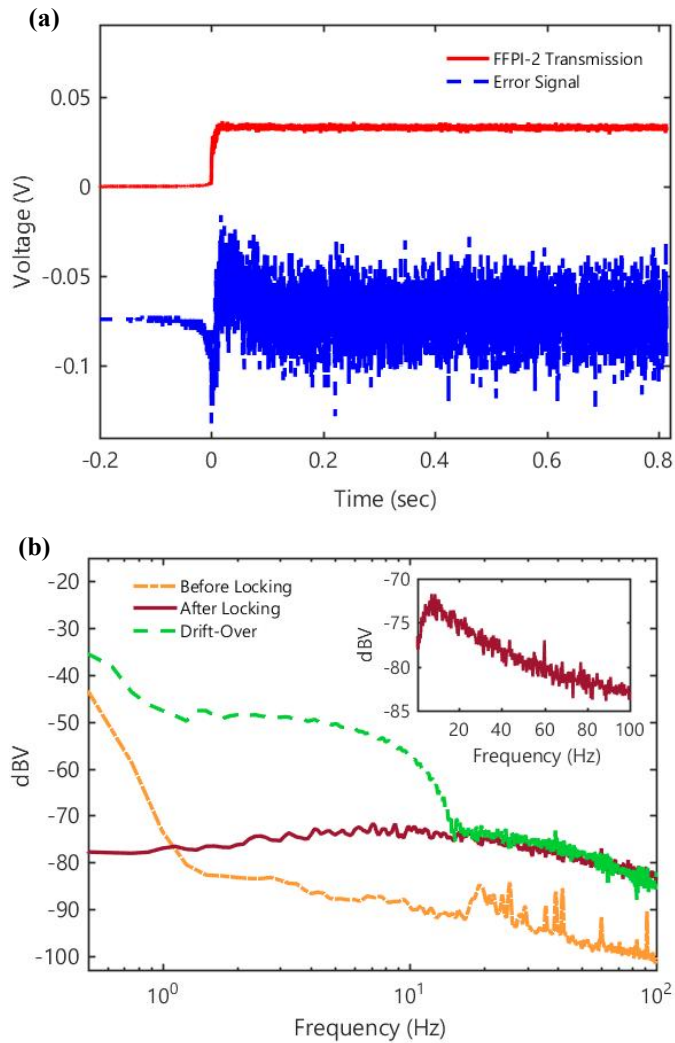

Fig. 2. Frequency locking between the diode laser and FFPI-2: (a) optical transmission of FFPI-2 (upper) and the PDH error signal (lower) before and after locking; (b) Fourier spectra of the error signal before locking (dash-dotted), after locking (solid), and under the "drift-over" condition (dashed).

### III. Experimental Result

To perform the experiment, the diode laser is first locked to the reference cavity (FFPI-2). Fig. 2 shows the behaviors of the PDH error signal before and after locking is established. In Fig. 2(a) (lower trace), the transition of the error signal from prior-locking to post-locking is captured in the time domain. At the moment of locking (time 0), the error signal experiences a rapid transient. After that, it quickly settles down to a steady level, albeit with significantly increased high-frequency fluctuations. The presence of a constant level of high-frequency noise in the error signal indicates that the laser frequency is tracking the slow drift of the resonance peak without responding to high-frequency jitters of the cavity. The establishment of laser-cavity locking is also evident from the behavior of optical transmission through FFPI-2, which is shown by the upper trace in Fig. 2(a). The photodetector output jumps from zero to a steady level at time 0 and remains on that level thereafter, indicating that a stable FFPI-2 transmission is established. This can only happen when the laser frequency and the cavity resonance are locked.

Meanwhile, it is also instructive to examine the behaviors of the error signal in the frequency domain, which are shown in Fig. 2(b). The dash-dotted (orange) trace on the bottom shows the Fourier spectrum of the error signal before locking. Since without locking the laser frequency is generally far away from any cavity resonances, the error signal is effectively zero and its spectrum is dominated by the instrument noise. Once the laser is locked to the cavity, the error signal displays a much-elevated level of noise, as shown by the solid (dark red) trace in Fig. 2(b). This is in agreement with the time-domain observation in Fig. 2(a). An interesting fact revealed by the frequency-domain measurement, however, is a decrease of the error signal below 10 Hz, which is a clear indication of the servo action. This downward bending of error signal spectrum at low frequencies can be more clearly seen on the linear frequency scale as shown in the inset of Fig. 2(b). To further showcase the effectiveness of our frequency locking system, we tune the laser frequency very close to a cavity resonance and measure the error signal spectrum while the resonance peak freely drifts over the free-run laser. The result is shown as the dashed (green) trace in Fig. 2(b). The high-frequency portion of the trace (>20 Hz) overlaps with the locked error signal spectrum, suggesting that they are of the same nature, i.e. dominated by fast jitter of the resonance peak. Below about 20 Hz, the free-run spectrum exhibits a substantial increase due to the relative frequency drift between the laser and the cavity resonance. Such a drift is completely removed once the frequency locking is engaged.

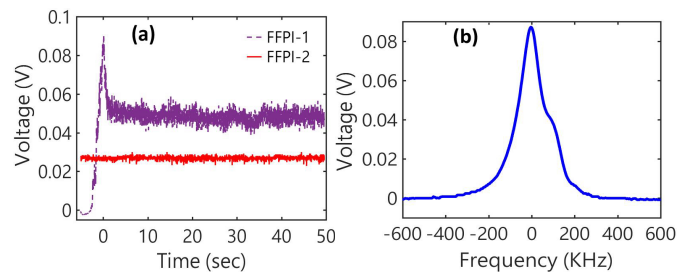

Fig. 3. (a) FFPI-1 (dashed) and FFPI-2 transmissions during the laser frequency tuning. The laser frequency is eventually parked in the middle of the trailing edge of an FFPI-1 resonance. (b) The high-resolution lineshape of an FFPI-1 resonance is obtained by scanning the relative frequency between the resonance peak and the interrogation laser.

After the laser is locked to the reference FFPI, we tune the laser frequency to match one of the resonance frequencies of the FFPI sensor (FFPI-1). This is done by applying a proper bias voltage across the PZT stretcher mounted on FFPI-2. As the length of the reference cavity is gradually adjusted, the servo automatically adjusts the laser frequency to keep it locked with the cavity resonance. The advantage of this method lies in the fact that the random drift of the two FFPIs has very little impact on the frequency detuning between the interrogation laser and the FFPI sensor, because the two FFPIs experience the same random drift due to their similar ambient conditions. This allows the tuning process to be done in a well-controlled fashion, as demonstrated in Fig. 3(a). The dashed (purple) trace shows the optical transmission through FFPI-1. As the laser frequency scans across a resonance peak, the transmitted power sketches out part of the cavity line shape. For optimum sensing performance, we choose to park the laser frequency on the edge of a resonance peak. This can be seen in Fig. 3(a) as the FFPI-1 transmission stays at a fix level half way down from the peak. Once the laser tuning is finished, the interrogation laser and the FFPI sensor are able to stably maintain their relative frequency for an extended period. This allows precise strain measurement to be carried out. In the meantime, the locking between the laser and FFPI-2 remains intact throughout this tuning process, as evident from the stable FFPI-2 transmission (solid red trace in Fig. 3(a)). The excellent relative frequency stability between the laser and FFPI-1 also enables an accurate measurement of the transmission line shape for FFPI-1. This is done by periodically scanning the laser frequency across an FFPI-1 resonance and monitoring the transmitted power. Fig. 3(b) shows the measured resonance line shape, whose FWHM agrees with the aforementioned cavity specifications. Such high-resolution characterization would be very difficult to accomplish with conventional spectrometers or tunable lasers.

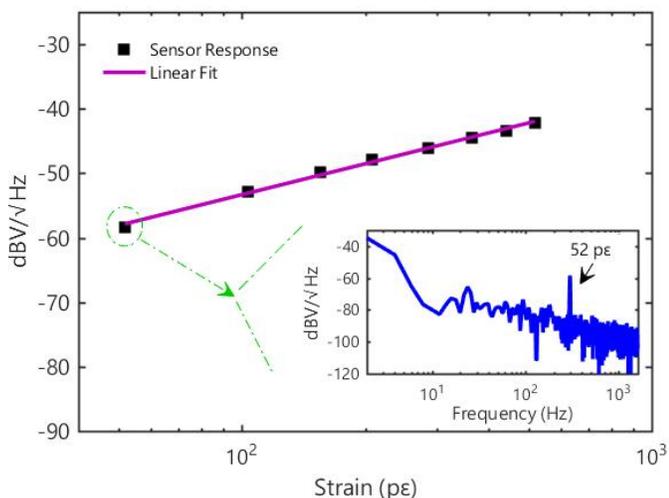

Fig. 4. FFPI sensor response to dynamic strains (at 300 Hz) of various amplitudes. Inset: Measured strain signal at 52 $p\varepsilon$ vs. noise floor

Measurement of *dynamic* strain is performed by applying a strain modulation to FFPI-1 (i.e. the FFPI sensor) while monitoring the variation of transmitted laser power. Strain modulations of various frequencies and amplitudes are introduced via the PZT in FFPI-1. The actual amounts of strain are calibrated based on the manufacturer-specified PZT response, which has also been independently verified in our experiment. A typical frequency response of the sensor under a single-tone strain modulation (300 Hz in this case) is shown in Fig. 4 inset. Despite a very small strain amplitude of 52 $p\varepsilon$, a signal-to-noise ratio (SNR) of 32 dB is obtained. Repeating such measurement with a number of strain modulation amplitudes leads to a strain-response curve, which is found to be highly linear as shown in Fig. 4.

The noise-limited strain resolution can be determined based on the above strain measurement and a measurement of the sensor noise floor. Fig. 5 shows the resulted strain resolution spectrum, along with the system noise background, over a Fourier frequency range of 10 mHz – 100 kHz. Apart from several spurious noise spikes, strain resolution is found to be equal or below 800 $f\varepsilon$ /√Hz at any frequency above 2 Hz. In particular, the sensor response stays fairly flat within 1–100 Hz with a nominal strain resolution of approximately 800 $f\varepsilon$ /√Hz. The noise spike near 30 Hz is likely due to mechanical or electrical coupling into the diode laser, which should be possible to remove with better isolation or shielding. Moving on to higher frequencies, the strain resolution reaches about 75 $f\varepsilon$ /√Hz at 1 kHz, 60 $f\varepsilon$ /√Hz at 2 kHz and 40 $f\varepsilon$ /√Hz at 23 kHz. These results are comparable to or even better than the strain resolutions achieved with phase-shifted FBGs [12] and slow-light FBGs [18]. They also appear to be better than some of the prior results based on FFPI sensors at frequencies above 1 kHz [34].

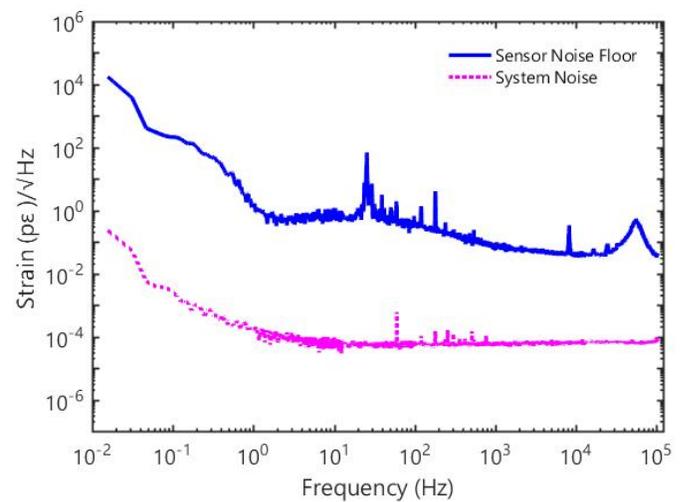

Fig. 5. Noise-limited strain resolution for the FFPI sensor over 7 decades of frequencies (10 mHz – 100 kHz). Lower trace shows the equivalent strain resolution due to the background system noise.

It should be pointed out here that the above ultrahigh strain resolutions have been achieved with only the basic *side-detection* scheme, using an *off-the-shelf* diode laser without any additional stabilization. The modest requirement in laser linewidth is especially interesting considering the fact that such scales of resolution typically require highly specialized narrow-linewidth lasers (e.g., random distributed fiber laser) or frequency referencing to OFC [12], [18], [34]. On the other hand, it is also well known that side-detection has some drawbacks compared to more sophisticated detection techniques such as the PDH method [25]. Chief among them is its sensitivity to laser intensity noise [38]. However, from a proof-of-concept point of view, the current work highlights the potential of using long, high-finesse FFPI as UHR sensors. In other words, high resolution can be achieved even with simple detection schemes. Future work will focus on improvement in the isolation system and the detection scheme.

## IV. CONCLUSION

In conclusion, direct frequency locking between a diode laser and a meter-long, high-finesse FFPI has been experimentally demonstrated. The laser assembly is used to interrogate a similar meter-long FFPI sensor and UHR dynamic-strain measurement has been carried out. A nominal, noise-limited strain resolution of 800 $f\varepsilon$ /√Hz has been achieved within 1–100 Hz. Strain resolution further improves to 75 $f\varepsilon$ /√Hz at 1 kHz, 60 $f\varepsilon$ /√Hz at 2 kHz and 40 $f\varepsilon$ /√Hz at 23 kHz, demonstrating a comparable or even better UHR potential than $\pi$-phase-shifted FBGs and slow-light FBGs. Since the current sensor resolution is mainly limited by laser intensity and frequency noise, additional measures in laser stabilization are expected to make further improvement in strain resolution. The work lays out a feasible path toward UHR fiber-optic sensing based on long FFPIs.